\documentclass[journal]{IEEEtran}
\usepackage[edges]{forest}
\usetikzlibrary{arrows.meta}
\usepackage{graphicx,times,amsmath,cite,enumerate}
\usepackage{kantlipsum}
\usepackage{threeparttable}
\usepackage{epsfig}
\usepackage{amssymb}
\usepackage{latexsym}
\usepackage{psfrag}
\usepackage{setspace}
\usepackage{color}
\usepackage{verbatim}
\usepackage{amsthm}
\usepackage{footnote}
\usepackage{textcase}
\usepackage{array}
\newcolumntype{P}[1]{>{\centering\arraybackslash}p{#1}}
\usepackage{float} 
\usepackage{multirow}
\usepackage{textcomp}
\usepackage{forest}
\PassOptionsToPackage{hyphens}{url}
\usepackage{url}
\usepackage{physics}
\usepackage{hyperref}
\usepackage{etoolbox}
\usepackage[edges]{forest}
\usepackage[ruled,vlined]{algorithm2e}
\usepackage{siunitx}

\usepackage{subfigure}
\usepackage{physics}
\usepackage{tikz}
\usetikzlibrary{automata, positioning}
\usepackage{nomencl}
\usepackage{float}
\usepackage{tabularx,booktabs}
\usepackage{footnote}

\renewcommand{\baselinestretch}{0.98}

\title{\vspace{-18pt}Decoupling Power Quality Issues in Grid-Microgrid Network Using Microgrid Building Blocks}

\IEEEoverridecommandlockouts

\begin{document}
\bstctlcite{IEEEexample:BSTcontrol} 


 \author{
    \IEEEauthorblockN{ Samrat Acharya}
    , \IEEEauthorblockN{ Priya Mana}
    , \IEEEauthorblockN{ Hisham Mahmood}
        , \IEEEauthorblockN{ Francis Tuffner}
        , \IEEEauthorblockN{ Alok Kumar Bharati}
    \\
    \IEEEauthorblockA{\textit{Pacific Northwest National Laboratory (PNNL), Richland, WA, USA}}\\

    \IEEEauthorblockA{(samrat.acharya, priya.mana, hisham.mahmood, francis.tuffner,  ak.bharati)@pnnl.gov \vspace{-8mm}}
}
\maketitle

\def\thefootnote{}\footnotetext{

The Pacific Northwest National Laboratory is operated by the Battelle Memorial Institute for the U.S. Department of Energy under contact DE-AC05-76RL01830. This work is funded by DOE Office of Electricity (OE).}
\def\thefootnote{\arabic{footnote}}

\begin{abstract}
Microgrids are evolving as promising options to enhance reliability of the connected transmission and distribution systems. 
Traditional design and deployment of microgrids require significant engineering analysis. Microgrid Building Blocks (MBB),  consisting of modular blocks that integrate seamlessly to form effective microgrids, is an enabling concept for faster and broader adoption of microgrids. Back-to-Back converter placed at the point of common coupling of microgrid is an integral part of the MBB. This paper presents applications of MBB to decouple power quality issues in grid-microgrid network serving 
power quality sensitive loads such as data centers, new grid-edge technologies such as vehicle-to-grid generation, and serving electric vehicle charging loads during evacuation before disaster events. Simulation results show that MBB effectively decouples the power quality issues across networks and helps maintain good power quality in the power quality sensitive network based on the operational scenario.

\end{abstract}
%
\IEEEpeerreviewmaketitle

\section{Introduction}
The power grid is evolving rapidly with increasing penetration of renewable in the system, especially in the lower voltage parts of the grid. The high penetration of Distributed Energy Resources (DERs) and increasing utility scale generation and storage is enabling increase in microgrid deployments \cite{deployment}. However, this broad microgrid deployment requires the power engineering community to enhance the functionality and features of a microgrid. 
Microgrid building blocks (MBB) is a similar concept that focuses on modularity for simplifying a microgrid design while making it powerful through advanced controls and communication  \cite{MBB_wp, chenching_ieeeaccess} . 

The MBB concept is inspired by the power electronics building blocks concept \cite{pebb}. Studies \cite{MBB_wp, chenching_ieeeaccess} provide details on the different types of MBB and the idea of utilizing them for developing a microgrid. These building blocks can be utilized to form a microgrid from scratch in greenfield scenarios or can be used in parts to aid existing grid components to enhance MBB functionality in brownfield scenarios. For example, if there is a smart inverter that has limited communication capability, adding a communication block can enhance the interaction of the smart inverter with rest of the microgrid components or controllers. 

The other key feature of the MBB concept is modularity and standardization \cite{MBB_wp}. The standardization is envisioned to be multi-fold: 1) Standard ratings; 2) Standard communication protocols; and 3) Standard terminals and interfaces \cite{MBB_wp}. These  features are expected to contribute significantly to the cost and quality of the MBB components due to economies of scale and competitive markets with multiple component manufacturers.

The US Department of Energy (DOE)'s Office of Electricity (OE) has been significantly  investing in the modular  microgrid concept and is currently driving research to make  it feasible  so that the  simplified microgrids that can be deployed with minimum engineering costs. Furthermore, DOE-OE's Microgrid R\&D Program is focused on enabling microgrid solutions for rural and undeserved communities, where MBB-based microgrids can be effective\cite{doe_microgrids}.

MBB concept enables a faster deployment of temporary microgrids due to the standard modular components that can be combined to serve a set of loads or support a weak system. There are specific applications and use cases that need fast deployment of microgrids such as: 1) Serving critical loads during and immediately after natural disasters such as hurricanes; 2) Deploying Electric Vehicle (EV) chargers (e.g., DC fast chargers) on evacuation routes in high EV penetration areas; 3) Serving tactical microgrids in military and relief operations; and 4)  Maintaining resource adequacy during planned maintenance of the grid that might result in loss of load in certain feeder sections. These applications can effectively utilize the MBB-based microgrids to ensure the load service with minimal disturbance to the rest of the system. 

Several previous works in the research community and industry on modular microgrids discuss plug and play microgrids \cite{modular1, modular2, modular3}. Furthermore, mobile microgrids have a similar approach to MBB \cite{mobile1, mobile2, mobile3}. However, mobile microgrids are designed for low power ratings, and hence, do not support large-scale applications such as energizing significant feeders serving MW-scale loads. Several studies utilize Back-to-Back (BTB) converters at the point of common coupling (e.g., \cite{majumder2009power, deng2021enhanced}). However, these studies conduct EMT simulations and have major challenges when it comes to scalability, complexity, and computational speed.

In this paper, we expand on aforementioned studies to show the functioning of MBB-based microgrids in specific operating conditions to decouple low power quality issues such as 3-phase voltage unbalance from rest of the connected systems. We use a dynamic model for the BTB converter that is developed and integrated into a distribution system solver, GridLAB-D, for system level simulations \cite{mahmood2024dynamic}. The use cases of the BTB converter are simulated on a grid-microgrid network, modified from the IEEE 13-node test system. The model considers 3-phase and 1-phase loads and DERs such as solar Photovoltaic (PV), Battery Energy Storage System (BESS), and EVs. We make following key contributions:

 \begin{table}[H]
\centering
\begin{tabular}{ |p{0.46\textwidth}| }
 \hline
\rule{0pt}{2.5ex}  This paper is accepted for publication in IEEE IECON 2024, Chicago, IL. The complete copyright version will be available on IEEE Xplore when the conference proceedings are published. \\
 \hline
\end{tabular}
\end{table}

\begin{enumerate}
    \item The paper presents results for system level 3-phase unbalanced phasor-domain dynamic simulations with a BTB converter and other dynamic components in a microgrid.
    
    \item We discuss practical applications of MBB-based microgrids applied to \textit{blue sky} and \textit{grey sky} days. Specifically, we scrutinize the operation of the BTB converter, which is an integral part of the MBB.

\end{enumerate}


\section{Building Blocks for MBB-based Microgrids}
\label{sec:blocks}
The main blocks in a MBB-based microgrid concept are power converter block; control block; communication block; and integrated block. These blocks are designed to operate and interface with each other in a seamless manner.

\subsection{Power Converter Block} 
The power converter block is a generalized block that contains all power conversion units, including rectifiers, inverters, and BTB converters that are aimed to interface several microgrid components such as generation (conventional such as diesel units, natural gas units and renewable such as solar and wind), load, and energy storage (mainly BESS). The power converter block has defined power and voltage ratings. 
PV inverters, BESS, UPS systems, and BTB converters are some examples of the power converter blocks.

\subsection{Control Block} 
This block is responsible for the control and coordination of the various microgrid components. The controllers are designed to enable optimal operation of microgrid and connected systems under blue sky scenarios with various objectives including,  to minimize greenhouse gas emitting generation (e.g., diesel, natural gas); to ensure sufficient headroom in energy storage resources to meet reserve during transients. Examples of this block are the microgrid controllers and MBB-based controller. The MBB-based control block is envisioned to enable advanced microgrid control and modular plug and play functionality.

\subsection{Communication Block}
The communication block is responsible for ensuring that the measurements and control signals are communicated across the microgrid components. The communication in a MBB-based microgrid is expected to be standardized. Moreover, the communication block is also expected to enable protocol translation to enable seamless connectivity between the various components of microgrids in brownfield and greenfield microgrids. Examples of communication block is a protocol converter that adds communication capability to inverters.

\subsection{Integrated Block}
The integrated block combines various functions of power conversion, control and communication blocks. The integrated block is typically located at the interface of the microgrid and power grid substation. The integrated block is usually capable of controlling and coordinating all the components of the microgrid including the performance of the power conversion block. The integrated block is deemed to be the most powerful among all the blocks.

Different combinations of these blocks can be made for greenfield and brownfield microgrids to enable advanced control and communication capabilities in the MBB-based microgrids. The paper explores the application of the integrated block with a BTB converter that has control and communication capability and manages the power import and export in the MBB-based microgrid.

\section{Case Study}
\label{sec:case_study}
\begin{figure}[!b]
    \centering
    \vspace{-20pt}
\includegraphics[width=0.99\columnwidth, clip=true, trim= 0mm 0mm 0mm 0mm]{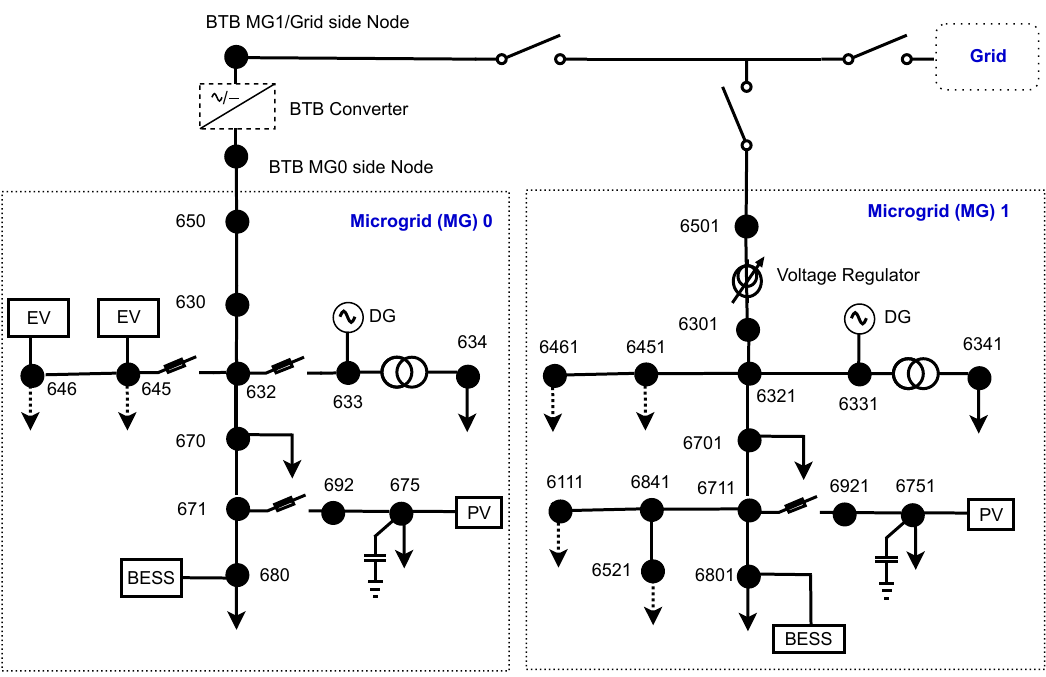}
    \caption{Test system with an MBB-based microgrid~(MG0) connected to the grid via a BTB converter at PCC, and a neighboring networked  microgrid~(MG1). Dashed lines with arrows represent single-phase loads and solid lines with arrows represent three-phase balanced loads.}
    \label{fig:test_system}
\end{figure}

As discussed in Section~\ref{sec:blocks}, BTB converter, a power conversion block, is an integral part of MBB.  After combining the power conversion block with the control and communication capabilities, the MBB can be implemented in versatile use cases. 
To help us discuss the use cases, we built a test system as shown in Fig.~\ref{fig:test_system}. The test system comprises two microgrids - MG0 and MG1, and the grid. MG0 is an MBB-based microgrid with DERs such as diesel generator (DG), BESS, PV, and constant power balanced three-phase and single-phase loads. MG1 is a networked microgrid without the MBB but with DERs such as  DG, BESS, PV, and constant power three-phase and single-phase loads. The switches between MG0, MG1, and the grid are configured for creating various configurations of simulation cases. Both MG0 and MG1 are modified IEEE 13-node systems with the addition of loads such as a data center, DERs, and EVs to integrate the changes in power systems.
MG1 has three-phase unbalanced loads as in the original IEEE 13-node feeder, whereas MG0 has three-phase balanced loads to depict data center/industrial loads/commercial EV charging stations in case~\ref{subsec:datacenter} and~\ref{subsec:MBB_ev_rich_comm}, and has single-phase aggregated loads to depict residential EVs and V2G functionality in case~\ref{subsec:v2g}.  

The system-level voltage is 4.16 kV except for node 634 in MG0 and node 6341 in MG1, which are stepped down by a 1 MVA, 4.16 kV/480 V transformer to 480 V.  A BESS is at node 680 in MG0 and node 6801 in MG1, a 1.6 MW PV unit is at node 675 in MG0 and node 6751 in MG1, and a 3 MVA diesel generator is at node 633 in MG0 and node 6331 in MG1. The BESS capacity in both MG0 and MG1 varies across the simulation cases below. The three-phase constant power balanced loads are distributed at nodes 634 and 6341 ($400$ kW and $290$ kVAR), 670 and 6701 ($200$ kW and $116$ kVAR), 675 and 6751 ($843$ kW and $462$ kVAR) and 680 and 6801 ($1155$ kW and $660$ kVAR) in MG0 and MG1 respectively. A shunt capacitor of 86.52 kVAR capacity is installed at nodes 675 in MG0 and 6751 in MG1 to help boost the node voltages. In MG1, the voltage regulator is modeled between nodes 6501 and 6301 to regulate the voltage based on the measurements at node 6801. Since MG0 is an MBB-based microgrid, a separate voltage regulator is not required because the MBB performs the voltage regulation in addition to other functions. Besides the balanced loads, single-phase loads in MG1 are distributed at nodes 6451 ($170$ kW and $125$ kVAR in phase B), 6461  ($230$ kW and $132$ kVAR in phase B), 6111  ($170$ kW and $80$ kVAR in phase C), and 6521  ($128$ kW and $86$ kVAR in phase A). We also have single-phase loads at nodes 645 ($680$ kW and $500$ kVAR  in phase B), 646  ($690$ kW and $396$ kVAR  in phase B) and battery systems (500 kVA each at nodes 645 and 646) in MG0. This combination of load and battery depict lumped residential loads with EVs for simulating V2G functionality.

The MBB is modeled using a BTB converter with additional controls and communication blocks and is located at the Point of Common Coupling (PCC) of MG0. We use the phasor domain model of the BTB converter, detailed in \cite{mahmood2024dynamic} for this study. The BTB converter has two AC/DC converters connected together by a DC-link. 
Both of the AC/DC converters in the BTB converter are rated as 3.5~MVA with the nominal DC-link voltage set to 8000~V. Among these two converters, one is responsible for regulating the DC-link voltage of the BTB converter (AC/DC Converter 1 at MG0-side), while the other is responsible for controlling the power flowing through the BTB converter (AC/DC Converter 2 at MG1-side). The DC-link is a implement as a capacitor and its dynamics are modeled as shown in the results corresponding to the DC-link voltage (vertical axis labeled as DC Voltage in figures). The AC/DC Converter 1 is modeled as a grid-following current source converter and the AC/DC Converter 2 can operate in either grid-forming or grid-following mode. The model captures the charging and discharging dynamics of the DC-link, which is an important aspect of the BTB converter model~\cite{mahmood2024dynamic}. 

The test system described in this paper and the phasor domain model of the BTB converter for MBB are implemented in the distribution system simulation tool GridLAB-D. GridLAB-D is an open-source simulation platform developed by the Pacific Northwest National Laboratory for modeling and simulating the operation and behavior of three-phase unbalanced distribution power systems in steady state, quasi-steady state, and dynamic timescales \cite{8307765,8267335}. Three different case studies for practical MBB-based applications are discussed next. These include data center operation, V2G generation, and EV charging for emergency evacuation.

\subsection{Modular Data Center Operation in MBB-based Microgrid}
\label{subsec:datacenter}

\begin{figure}[!htb]
\centering
\vspace{-1mm}
\subfigure[\label{fig:load_mg1_vuf}]{\includegraphics[width=0.9\columnwidth, clip=true, trim= 2.5mm 7mm 2.5mm 2mm]{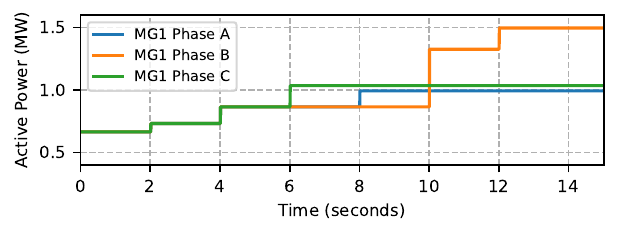}}

\subfigure[\label{fig:vuf_vuf}]{
\includegraphics[width=0.9\columnwidth, clip=true, trim= 2.5mm 7mm 2.5mm 2mm]{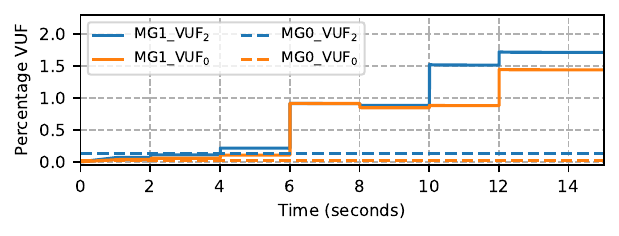}}

\subfigure[\label{fig:btb_P_vuf}]{
\includegraphics[width=0.9\columnwidth, clip=true, trim= 2.5mm 7mm 2.5mm 2mm]{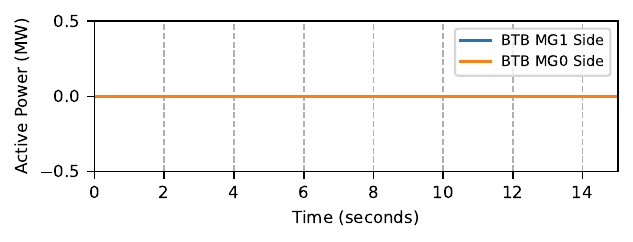}}

\subfigure[\label{fig:btb_Vdc_vuf}]{
\includegraphics[width=0.9\columnwidth, clip=true, trim= 2.5mm 7mm 2.5mm 2mm]{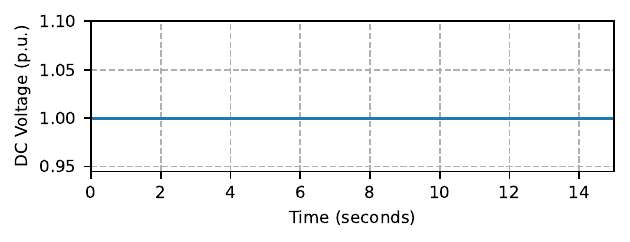}}


\subfigure[\label{fig:bess_mg1_vuf}]{
\includegraphics[width=0.9\columnwidth, clip=true, trim= 2.5mm 7mm 2.5mm 2mm]{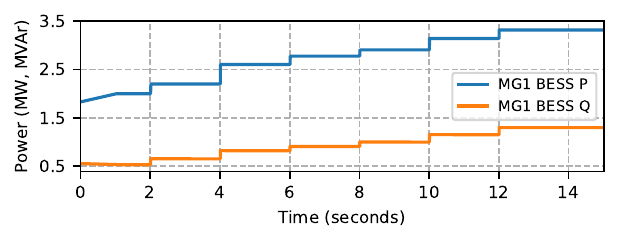}}

\subfigure[\label{fig:freq_vuf}]{\includegraphics[width=0.9\columnwidth, clip=true, trim= 2.5mm 7mm 2.5mm 2mm]{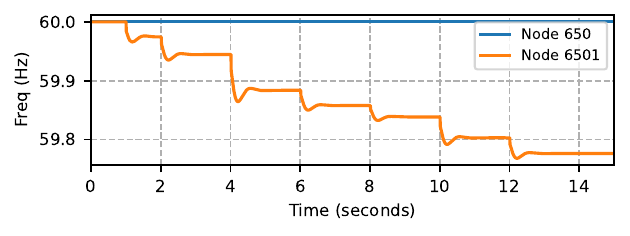}}

\subfigure[\label{fig:voltage_vuf}]{\includegraphics[width=0.9\columnwidth, clip=true, trim= 2.5mm 2mm 2.5mm 2mm]{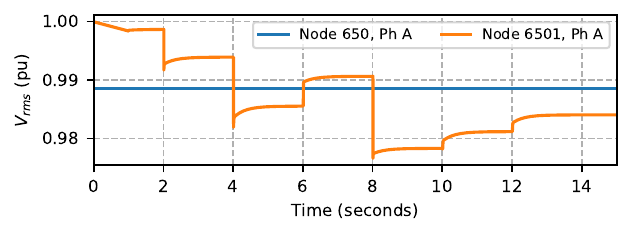}}

\label{fig:vuf}

\vspace{-2.5mm}
\caption{Voltage unbalance from external feeders does not propagate to MBB-based microgrid (MG0) with power-sensitive data center load. a) Phase-wise unbalanced load switching in MG1, b)  percentage voltage unbalance factor in MG0 and MG1, c) power transfer in microgrids via BTB converter, d) BTB converter DC-link voltage, e) MG1 BESS power supply to meet MG1 loads, f) frequencies at MG0 and MG1, and g) RMS voltage at Nodes 650 and 6501. }
\vspace{-3.5mm}
\end{figure}

With the massive adoption \& advancement of artificial intelligence \& data management needs, there is a steep growth in the demand for data centers. To meet this demand, in addition to newer facilities being developed, urban cooperate buildings, warehouses and other facilities are getting converted to data centers~\cite{data_centers}. Using the existing grid infrastructure at the edge to deliver power to such data centers might incur power quality concerns as the data centers require high power quality and reliability. One such power quality issue can be voltage unbalances of neighboring feeders impacting the feeder with the data center load. It can be challenging to isolate the data centers from the voltage unbalance issues in the rest of the grid. Furthermore, having dedicated feeders for data centers might not be feasible or cost-effective. 
This case discusses the use of  MBB to provide isolation from voltage unbalances and maintain reasonable power quality in the data centers.

To explore this case study, a data center load is added to the test system in Fig.~\ref{fig:test_system}, which simulates an urban grid network with connected radial feeders. We model MG1 as an unbalanced microgrid and MG0 as a three-phase balanced microgrid serving data center loads. To simulate this case study, we open the switch between nodes 632 and 645 to remove the unbalanced lateral. To replicate the data center load, the load at node 634 in MG0 is modeled as 1236 kW, a massive step up from the 400 kW load that was originally assigned to node 634 in the IEEE 13-node feeder to simulate a high demand of data center load served by a distribution transformer of 4.16 kV/480V. The data center load is derived from the benchmarking study presented in~\cite{data_center_benchmarking}.


To show a high unbalance scenario, we disconnect the two microgrids from the grid to create a set of networked microgrids. The simulated experiment involves increasing the voltage unbalance in the neighboring microgrid, MG1, to study the impact on performance of the MBB based microgrid, MG0, and evaluate the MBB-based solution to isolate the voltage unbalances. The voltage unbalance is simulated by turning on unbalanced loads in the neighboring microgrid, MG1. Here, we define Voltage Unbalance Factor (VUF) based on negative and zero sequence components as:
\begin{equation}
    VUF_2 = \frac{\lvert V_2\lvert}{\lvert V_1 \lvert}* 100\%, \quad VUF_0 = \frac{\lvert V_0\lvert}{\lvert V_1 \lvert}* 100\%,
\end{equation}

where $\lvert V_0\lvert$,  $\lvert V_1\lvert$, and $\lvert V_2\lvert$ are magnitude of zero, positive, and negative sequence components of the three-phase voltage.

Fig.~\ref{fig:load_mg1_vuf} shows the aggregated phase-wise loads in MG1. We start the simulation with three-phase loads at nodes 6751 and 6801 served in MG1 with its BESS. The microgrid MG0 is operating at its rated load (data center loads and other loads) with power supplied from DG, PV, and BESS. In MG1, at $t=2$ and $t=4$ seconds, we bring three-phase loads at nodes 6701 and 6341 online, respectively.

\begin{figure}[!htb]
\centering
\vspace{-0mm}

\subfigure[\label{fig:btb_P_v2g}]{
\includegraphics[width=0.94\columnwidth, clip=true, trim= 2.5mm 7mm 2.5mm 2.5mm]{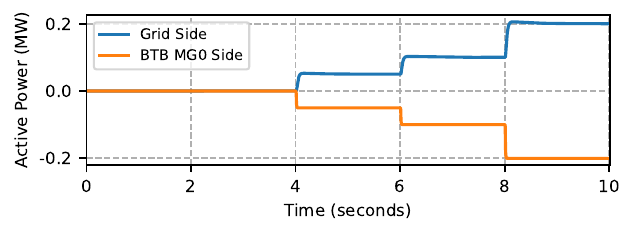}}

\subfigure[\label{fig:btb_Vdc_v2g}]{
\includegraphics[width=0.94\columnwidth, clip=true, trim= 2.5mm 7mm 2.5mm 2.5mm]{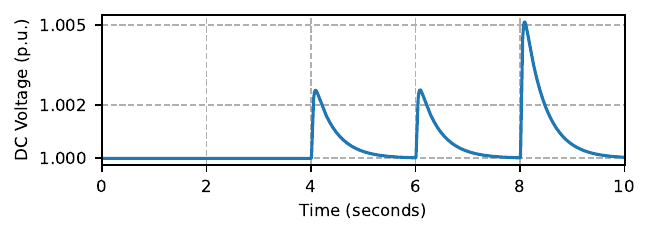}}

\subfigure[\label{fig:bess_unbalance_v2g}]{
\includegraphics[width=0.94\columnwidth, clip=true, trim= 2.5mm 7mm 2.5mm 2.5mm]{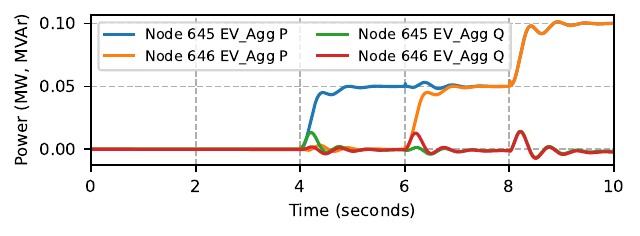}}

\subfigure[\label{fig:bess_680_v2g}]{
\includegraphics[width=0.94\columnwidth, clip=true, trim= 2.5mm 7mm 2.5mm 2.5mm]{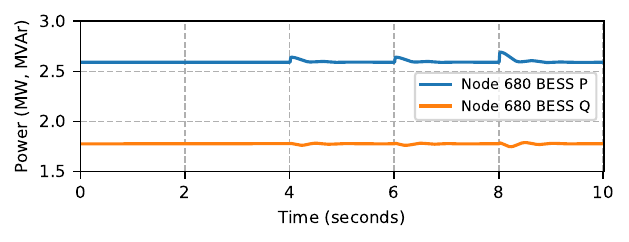}}


\subfigure[\label{fig:freq_v2g}]{\includegraphics[width=0.92\columnwidth, clip=true, trim= 2.5mm 7mm 2.5mm 2.5mm]{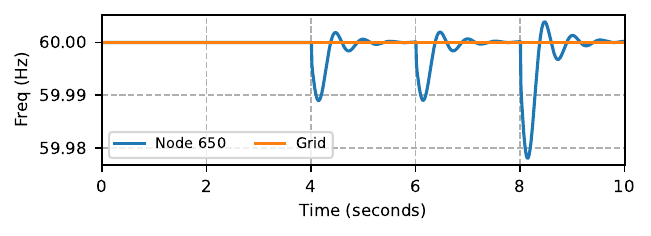}}

\subfigure[\label{fig:vuf_v2g}]{\includegraphics[width=0.92\columnwidth, clip=true, trim= 2.5mm 7mm 2.5mm 2.5mm]{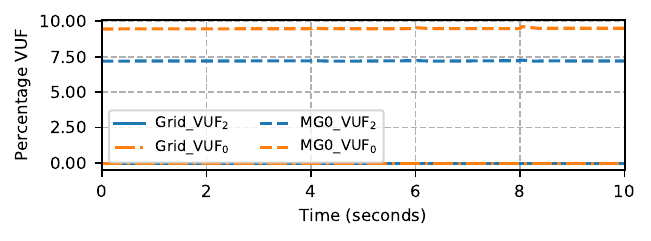}}

\subfigure[\label{fig:power_phase_v2g}]{
\includegraphics[width=0.94\columnwidth, clip=true, trim= 2.5mm 2mm 2.5mm 2.5mm]{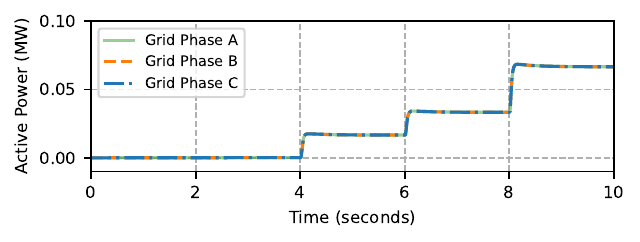}}

\label{fig:v2g}

\vspace{-2mm}
\caption{MBB-based microgrid (MG0) operating with unbalanced V2G power generation but supplying balanced power to grid due to isolation provided by the BTB converter. a) Power transferred to grid from MG0 via BTB converter, b) DC-link voltage of the BTB converter, c) V2G BESS power at nodes 645 and 646, d) BESS power at node 680, e) frequency, f) VUF, and g) Phase wise power transferred to grid via BTB.}
\vspace{-3.5mm}
\end{figure}

Before $t=6$ seconds, $VUF_0$ and  $VUF_2$ in Fig.~\ref{fig:vuf_vuf} are zero in MG0 while it is $\approx 0.1\%$ in MG1, which is due to the unbalanced system configuration in MG1. At $t=6$, $t=8$, $t=10$, and $t=12$ seconds, we bring single phase loads at nodes 6111, 6521, 6461, and 6521 online, respectively. At $t=6$ second, due to turning on phase C load, both VUFs in MG1 increased to $\approx 1\%$. 
However, at $t=8$ second, when phase A load at node 6521 is turned on, both VUFs slightly decreased as phase loads became more balanced than at $t=6$ seconds. Similarly, when phase B loads are turned on at $t=10$ and $t=12$ seconds, $VUF_2$ and $VUF_0$ increase to $\approx 1.8\%$ and $\approx 1.48\%$, respectively. Notably, during these single-phase loads turning on at MG1, both VUFs at MG0 remains unaffected. Furthermore, the BTB converter does not exchange power between MG0 and MG1 as in Fig.~\ref{fig:btb_P_vuf}. The DC-link voltage is 1 per unit through the simulation as in Fig.~\ref{fig:btb_Vdc_vuf}. The loads in MG1 is entirely supplied by its BESS as in Fig.~\ref{fig:bess_mg1_vuf}. As the MG1 BESS is in grid forming mode, the frequency of MG1 is set to new operating point for every load change in MG1 as in Fig.~\ref{fig:freq_vuf}. However, the frequency at MG0 remains at 60 Hz. Similarly, three-phase rms voltage at MG0 is constant at 1 per unit as in Fig.~\ref{fig:voltage_vuf}, while it is varying at MG1 although with a small magnitude, which depends upon the system strength for voltage regulation. These simulation results provide a preliminary results to operate data centers as a part of an MBB-based microgrid.


\subsection{Vehicle-to-Grid (V2G) Operation in MBB-based Microgrid}
\label{subsec:v2g}
This use case discusses blue-sky operations where a feeder with multiple single-phase residential loads is modeled. With rapid EV adoption among customers, it is assumed that some of the single-phase residential loads have EVs plugged into residential circuits which are capable of providing V2G capability. It is possible that the single-phase V2G capability creates an unbalanced generation at the grid-edge. In such scenarios, an MBB-based microgrid can provide isolation to the microgrid, so that the microgrid on the whole exports balanced power to the grid. 

To adapt this scenario to the test system discussed in this paper, we assume aggregated single-phase residential loads with V2G capability. To simulate this case, we configure the test system in Fig.~\ref{fig:test_system} to connect the MBB-based microgrid (MG0) to the grid via BTB converter. We open switches between nodes 632 and 633 and nodes 671 and 692 in MG0, and disconnect load at node 670. Thus, we have V2G responses from three nodes, i.e., 680, 645, and 646. The loads at nodes 645 and 646 are constant power phase B loads, while the load at node 680 is a three-phase constant power load.

As the simulation progresses, the loads at nodes 645 and 646 are increased to 690 kW  and 396 kVAR and 680 kW  and 500 kVAR, respectively, which adds to the voltage unbalance in the system. We control BTB converter to supply 50 kW, 100 kW, and 200 kW power from MG0 to grid at $t=4$, $t=6$, and $t=8$ seconds, respectively as shown in Fig.~\ref{fig:btb_P_v2g}. The DC-link voltage of the BTB converter is well regulated to 1 per unit during these power transfers as in Fig.~\ref{fig:btb_Vdc_v2g}. The power transferred to the grid is generated by an aggregation of V2G enabled EV BESS at nodes 645 and 646 as shown in Fig.~\ref{fig:bess_unbalance_v2g}.
At $t=4$ second, EV BESS at node 645 supplies 50 kW, at $t=6$ second, EV BESS at 646 supplies 50 kW, and at $t=8$ seconds, both EV BESS supply 50 kW each to meet the power transferred to the grid via BTB converter. In these power transfers, BESS at node 680 in Fig.~\ref{fig:bess_680_v2g} is in grid forming mode, and hence, regulates frequency as shown in Fig.~\ref{fig:freq_v2g}. Since the load at nodes 645 and 646 are in phase B, $VUF_2$ and $VUF_0$ in MG0 are above 7\% and 9.8\%, which are higher than the practiced limit of 2\% as shown in Fig.~\ref{fig:vuf_v2g}. However, both VUFs at grid are almost zero. Similarly, the phase power in grid are balanced in Fig.~\ref{fig:power_phase_v2g}, which means despite the low power quality in MG0 due to unbalanced loads and V2G responses, we can export balanced power to grid or connected microgrid via BTB converter using the MBB components.
\begin{figure}[!htb]
\centering
\subfigure[\label{fig:load_ev_evacuation_full}]{\includegraphics[width=0.97\columnwidth, clip=true, trim= 2.5mm 7mm 2.5mm 2.5mm]{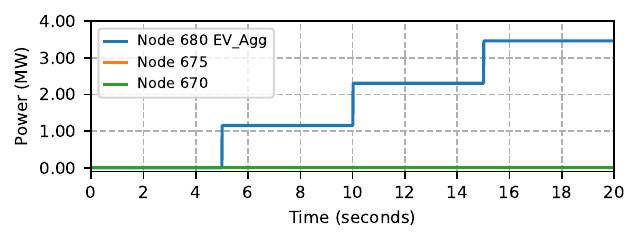}}

\subfigure[\label{fig:btb_P_ev_evacuation_full}]{
\includegraphics[width=0.97\columnwidth, clip=true, trim= 2.5mm 7mm 2.5mm 2.5mm]{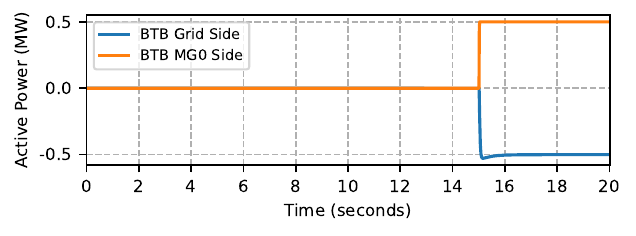}}

\subfigure[\label{fig:btb_Vdc_ev_evacuation_full}]{
\includegraphics[width=0.97\columnwidth, clip=true, trim= 2.5mm 7mm 2.5mm 2.5mm]{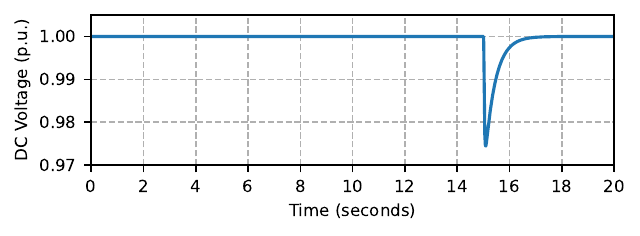}}

\subfigure[\label{fig:bess_ev_evacuation_full}]{
\includegraphics[width=0.97\columnwidth, clip=true, trim= 2.5mm 7mm 2.5mm 2.5mm]{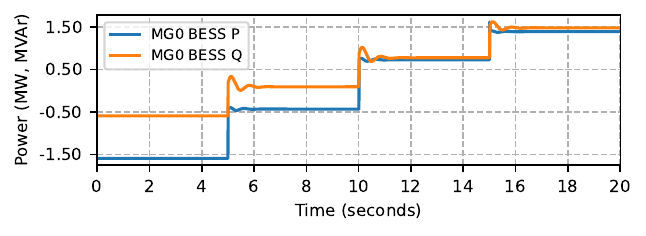}}

\subfigure[\label{fig:pv_ev_evacuation_full}]{
\includegraphics[width=0.97\columnwidth, clip=true, trim= 2.5mm 7mm 2.5mm 2.5mm]{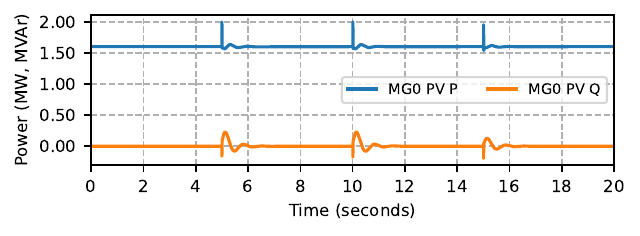}}

\subfigure[\label{fig:freq_ev_evacuation_full}]{\includegraphics[width=0.97\columnwidth, clip=true, trim= 2.5mm 7mm 2.5mm 2.5mm]{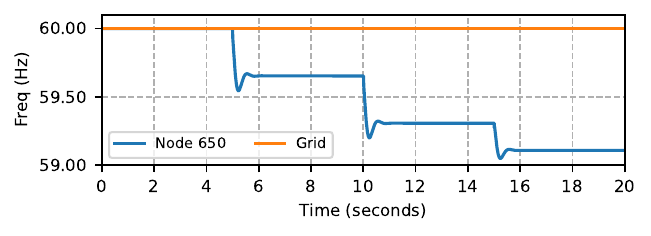}}

\subfigure[\label{fig:voltage_ev_evacuation_full}]{\includegraphics[width=0.97\columnwidth, clip=true, trim= 2.5mm 7mm 2.5mm 2.5mm]{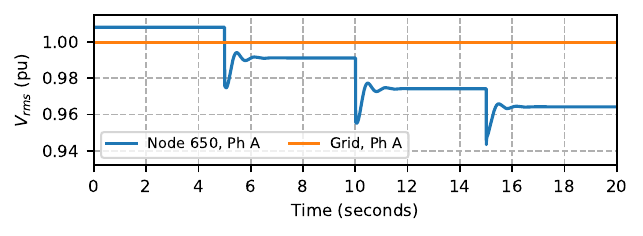}}


\label{fig:ev_evacuation}
\vspace{-2mm}
\caption{MBB-based microgrid serving EV loads during evacuation while isolating the grid from poor power quality. a) EV charging load at bus 680 in MG0, b) power exchange between MG0 and grid via BTB converter, c) BTB DC-link voltage, d) MG0 BESS power, e) MG0 PV power, f) frequencies at MG0 and grid, g) Phase A RMS voltages at MG0 and grid.}
\vspace{-3.5mm}
\end{figure}

\subsection{MBB-based Microgrids Operations under Disaster Events}
\label{subsec:MBB_ev_rich_comm}
While maintaining power quality is important during blue-sky and normal operations, in case of a disaster event such as approaching wildfires or an incoming hurricane, the priority shifts to safe grid operations, while also being able to provide power for critical functions. With the increase in EV adoption, a large population relies on EV charging facilities for readying them for evacuation and providing them mobility as needed. An MBB-based microgrid can provide a solution in such scenarios where critical loads such as vehicle charging stations can be kept operational, albeit at a lower power quality, to ensure the customers are not left without any power. Additionally, it is best if a poor power quality microgrid is isolated, even if electrically connected to the grid, so that the grid which is possibly already strained during the disaster event, doesn't collapse.  This section presents the ability of an MBB-based microgrid (MG0) to supply its critical loads at lower power quality during emergencies while being connected to the grid. We consider the case of EV rich community, where EV charging is critical before evacuating the area. To demonstrate this case, we configure test system in Fig.~\ref{fig:test_system} to MG0 connected to grid via BTB converter. Furthermore, we open switches between nodes 632 and 633 and 632 and 645 and disconnect loads at nodes 670 and 675. This configuration in MG0 means we only supply high EV charging demand at node 680 with BESS at node 680 and PV at node 675. 

We simulate a situation with a rapid increase in EV charging demand as customers prepare to evacuate an area. As Fig.~\ref{fig:load_ev_evacuation_full} shows, EV charging load at node 680 is zero till $t=5$ seconds. At $t=5$ seconds, EV charging load increases to 1155 kW, which is equivalent to 23 EVs charging at 50 kW high-power DC chargers. At $t=10$ seconds, we double the charging demand. Similarly, at $t=15$ seconds, the EV charging demand becomes 3465 kW, equivalent to 70 EVs charging at the DC chargers. During this emergency situation, MG0 fully supplies the EV charging load on its own till $t=15$ seconds. However, MG0 generation becomes insufficient at $t=15$ seconds, and hence, draws power from the grid via BTB converter as shown in Fig.~\ref{fig:btb_P_ev_evacuation_full}. The DC-link voltage of the BTB converter during this power transfer is regulated back to 1 per unit by the BTB controller as shown in Fig.~\ref{fig:btb_Vdc_ev_evacuation_full}. Fig.~\ref{fig:bess_ev_evacuation_full} shows the BESS power which is charging before $t=5$ seconds from the PV power in Fig.~\ref{fig:pv_ev_evacuation_full}, and supplies the EV charging power after $t=5$ seconds due to the increased load demand in the microgrid. During this emergency situation, frequency at MG0 falls to 59.1 Hz, while the grid frequency remains at 60 Hz as shown in Fig.~\ref{fig:freq_ev_evacuation_full}. Similarly, phase A voltage measured at node 650 terminal, which is also the voltage at the MG0 terminal falls to 0.946 per unit, which is significantly lower than the phase A grid voltage as shown in Fig.~\ref{fig:voltage_ev_evacuation_full}. We show only phase A voltage at MG0 and grid because all phase voltages are almost balanced in this simulation. 
These frequency and voltage profiles at MG0 show the capability of the BTB converter to allow the MBB-based microgrid to remain connected to the grid and exchange power during emergencies irrespective of the power quality at MG0 and grid during emergencies.






\section{Conclusion}
\label{sec:conclusion}
  Modular microgrids concept has been discussed in the literature and the industry has shown increased interest in it. While the concept seems promising, there is little work done to study the details of how such a system would operate. MBB as a concept is being explored and has seen interest from federal agencies for broad applications and widespread adoption of microgrids. In this paper, the BTB converter is explored as an important MBB component. This paper discusses three practical applications with different DER mixes and operating conditions of an MBB-based microgrid.

  There are two cases of power quality isolation that are discussed that show that MBB-based microgrids can import or export power without compromising the power quality on the sides where it is important. The data center application discussed has the data center on the microgrid and is sensitive to power quality. The MBB at the point of common coupling helps to import power from an unbalanced grid but ensures it is balanced in the microgrid. Similarly, in the V2G use case, the MBB-based microgrid exports power to the grid for maximum resource utilization. While the microgrid is unbalanced due to unbalanced power generation at the grid edge, the power exported out to the grid is balanced as preferred by operators.

  Furthermore, we present a disaster response study. To serve maximum EV charging load during an emergency, the MBB-based microgird in an EV rich community operates at relatively low frequencies and poor power quality. The MBB isolates power quality issues in the microgrid operating in emergency mode from rest of the grid to reduce strain on the grid and maintain nominal frequency on the grid-side. 
  
  These use case simulations demonstrate the benefits of MBB-based microgrids and the concept itself is effective. While there are advances being made to develop models for modular microgrids for simulation case studies, there is need for future standards and detailed cost-benefit analysis to enable a wide spread adoption of  modular microgrid technologies.

 

\bibliographystyle{IEEEtran}
\bibliography{MBB2_ref}
\end{document}